\documentclass[11pt,a4paper]{article}

\usepackage{textcomp}
\usepackage{amssymb}
\usepackage{amsmath}

\usepackage{graphicx}
\usepackage{latexsym}
\usepackage{appendix}
\usepackage{srcltx}
\def\CO{{\cal O}}

\def\CW{{\cal W}}

\def\centeron#1#2{{\setbox0=\hbox{#1}\setbox1=\hbox{#2}\ifdim
   \wd1>\wd0\kern.48\wd1\kern-.48\wd0\fi
   \copy0\kern-.48\wd0\kern-.48\wd1\copy1\ifdim\wd0>\wd1
   \kern.48\wd0\kern-.48\wd1\fi}}

\def\PRL{Phys. Rev. Lett.~}
\def\PR {Phys. Rev.~}

\newcommand{\beq}{\begin{equation}}
\newcommand{\eeq}{\end{equation}}
\newcommand{\bea}{\begin{eqnarray}}
\newcommand{\eea}{\end{eqnarray}}
\newcommand{\ba}{\begin{array}}
\newcommand{\ea}{\end{array}}

\newcommand{\p}{\partial}
\newcommand{\nn}{\nonumber}

\newcommand{\half}{\frac{1}{2}}

\renewcommand{\thefootnote}{\arabic{footnote}}

\begin{document}

\hskip11.6cm{CQUeST--2012-0559}
\vskip1cm

\begin{center}
 \LARGE \bf
Fake Supersymmetry  and Extremal Black Holes
\end{center}

\vskip1.5cm

\centerline{\large Seungjoon Hyun${}^{1a}$, ~
 Jaehoon Jeong${}^{2b}$, ~  Sang-Heon
Yi${}^{2c}$}

\hskip1cm

\begin{quote}{${}^{1}$Department of Physics, College of Science, Yonsei University, Seoul 120-749, Korea \\
${}^{2}$Center for Quantum Spacetime, Sogang University, Seoul 121-741,
Korea}\\
\end{quote}
\hskip2cm

\centerline{\large \bf Abstract}
\hskip0.3cm
We derive  the BPS type of first order differential equations for the rotating black hole solutions in the three-dimensional Einstein gravity coupled minimally with a self-interacting scalar field, using fake supersymmetry formalism. It turns out that the formalism is not complete and should be augmented by an additional equation to imply the full equations of motion. We identify this additional equation as a constraint by using an effective action method. By computing the renormalized boundary stress tensor, we obtain the mass and  angular momentum of the black hole solutions of these first order equations and confirm that they saturate the BPS bound.

\hskip2cm

\noindent\underline{\hskip12cm}\\
 {\it \small  ${}^{a}$sjhyun(at)yonsei.ac.kr~  ${}^{b}$jhjeong(at)sogang.ac.kr~
 ${}^{c}$shyi(at)sogang.ac.kr}

\thispagestyle{empty}
\renewcommand{\thefootnote}{\arabic{footnote}}
\setcounter{footnote}{0}

\newpage


\section{Introduction}
Supersymmetry, if confirmed experimentally, has a profound significance in our nature. It would give us  various predictions and new perspectives for particle phenomenology and cosmology. Apart from these implications, it explains systematically many interesting analytic results which may be ad-hoc or difficult to understand otherwise. This analytic nature is more or less related to the so-called Bogomol'nyi-Prasad-Sommerfield (BPS) states in extended supersymmetric field theories, which preserve  supersymmetry partially. One interesting and nice aspect of these BPS states is that they admit the Killing spinors which satisfy the Killing spinor equations (KSE). These KSE are usually lower order differential equations than the original equations of motion and therefore are easier to solve. Explicitly,   in the usual two derivative theory, the bosonic  equations of motion are given by second order differential equations, while   BPS states can be described by first order equations.

Typically these BPS states exist even in the model which contains only bosonic sector of the supersymmetric theory. In this reduced model they usually correspond to the states which minimize the energy, from which, once again, the lower order equations can be obtained.  Inspired by this,  fake supersymmetry method  has been developed to  obtain these BPS states which satisfy lower order equations of motion for non-supersymmetric, i.e. purely bosonic,
model \cite{Skenderis:1999mm}\cite{DeWolfe:1999cp}\cite{Freedman:2003ax}. The basic idea  is simple: One may consider a `fake' supersymmetric extension of the bosonic model and introduce a  spinor which satisfies fake KSE in the corresponding supersymmetric model. Since the EOM of original bosonic model are the same as the bosonic EOM of the supersymmetric model for vanishing fermions,  these reduced order `fake' KSE would almost imply the full EOM as in genuine supersymmetric theory.    

Along this line, various interesting BPS solutions of 
 gravity
with a
minimally coupled scalar field have been found. They include domain wall solutions \cite{Skenderis:1999mm} and  black hole solutions with a scalar hair \cite{Ceresole:2007wx}. They were found by considering some reduced EOM which are consistent with the full EOM. In the case of domain wall solutions and some static black hole solutions,  those reduced EOM have been obtained by using fake supersymmetry formalism \cite{Freedman:2003ax}.

In this paper we would like to establish a systematic method to obtain these reduced order EOM    by using fake supersymmetry formalism. 
Specifically, we consider the three-dimensional Einstein gravity
with a
minimally coupled and self-interacting scalar field. It would be considered as a bosonic sector of some fake supergravity. It was found that the model admits asymptotically anti-de Sitter black hole solutions with a scalar hair \cite{Henneaux:2002wm}\cite{Martinez:2004nb}\cite{Kwon:2012zh}\cite{Hotta:2008xt} as well as Banados-Teitelboim-Zanelli (BTZ) black holes \cite{Banados:1992wn}. 
We use the KSE of the fake supergravity to find the lower order EOM. It turns out that the KSE are not enough to uniquely determine the solutions. We find that it is due to the fact that the Killing vector associated with the fake Killing spinor is null-like. We identify the missing equation and argue that this corresponds to the constraint equation  rather than the dynamical EOM. In order to support the claim, we consider the effective action formalism. The resultant solutions are shown to correspond to quarter BPS solutions in the supersymmetric counter part. 

Since the solutions are asymptotically anti-de Sitter,  these can be studied in the context of AdS/CFT correspondence.  We determine the counter terms for the scalar and the metric fields and  compute renormalized boundary stress tensor. From this we 
obtain the mass and angular momentum of the solutions and confirm that they really saturate the BPS bound.

\section{Einstein gravity with an interacting scalar field}

The action of three-dimensional Einstein gravity with a minimally coupled scalar field is given
by
\begin{equation}\label{action}
S =\frac{1}{16\pi G}\int d^3x\sqrt{-g}\bigg[ R -\half
\p_{\mu}\phi \p^{\mu}\phi -V(\phi) \bigg]\,,
\end{equation}
where  we have taken the convention of the metric as  mostly plus signs and  the 
curvature tensors  as $[\nabla_{\mu}\, \nabla_{\nu}
]V_{\rho}
= R_{\mu\nu\rho\sigma}V^{\sigma}$ and  $R_{\mu\nu} =
g^{\alpha\beta}R_{\alpha\mu\beta\nu}$.

The EOM are composed of scalar field equation and the metric field equations  as follows ;
\beq 0 = E_{\phi} \equiv  \nabla^2\phi - \frac{\p V}{\p \phi}\,, \qquad
0={\bf E}_{\mu\nu} \equiv G_{\mu\nu} - T_{\mu\nu} \,,
 \eeq
where
\[ G_{\mu\nu} \equiv R_{\mu\nu} -\half R g_{\mu\nu} \,, \qquad T_{\mu\nu} \equiv   \half \p_{\mu}\phi \p_{\nu}
\phi  -\half g_{\mu\nu} \Big[ \half\p_{\alpha}\phi\p^{\alpha}\phi +   V(\phi)\Big]\,.  \]\
As usual, the trace part of ${\bf E}_{\mu\nu}$ can be used to rewrite EOM as
\beq 0 = E_{\mu\nu} \equiv  R_{\mu\nu} - \frac{1}{2}\p_{\mu}\phi\p_{\nu}\phi - g_{\mu\nu}V\,, \eeq
which is the relevant form for our study in next sections.

We are interested  in the asymptotically AdS black holes with a scalar hair, which would be  deformations of BTZ black holes.
Our metric ansatz for rotating AdS black holes with axial symmetry  in AdS-Schwarzschild-like coordinates  is taken  as
\bea
ds^2 &=& L^2 \Big[ -e^{2A(r)} dt^2 + e^{2B(r)}dr^2
+r^2\Big(d\theta + e^{C(r)}dt\Big)^2
 \Big]\,,
\label{BTZ}
\eea
where $L$ denotes the radius of asymptotic AdS space.   Accordingly, the scalar field $\phi$ is   taken as a function only of the radial coordinate $r$. The asymptotic conditions on the metric functions $A(r),B(r),C(r)$
are taken  as
\beq e^{A(r)}\Big|_{r\rightarrow\infty} \rightarrow r + \CO\Big(\frac{1}{r}\Big)\,, \qquad
e^{B(r)}\Big|_{r\rightarrow\infty} \rightarrow \frac{1}{r} + \CO\Big(\frac{1}{r^3}\Big)\,,
\qquad e^{C(r)}\Big|_{r\rightarrow\infty } \rightarrow const. +
\CO\Big(\frac{1}{r^2}\Big)\,. \eeq
The boundary condition for the scalar field consistent with this metric ansatz  is  given by
\beq \phi(r)\Big|_{r\rightarrow\infty} = const. + \CO\Big(\frac{1}{r}\Big)\,. \eeq
We have taken our  fall-off boundary conditions for the metric as the standard Brown-Henneaux type which
allow us to obtain the central charge  by
Brown-Henneaux method~\cite{Brown:1986nw}. One may   note  that the above metric admits a time-like Killing vector $\frac{\partial}{\partial t}$ and a rotational Killing vector $ \frac{\partial}{\partial \theta}$, which would generate the full isometry group in the generic case as was shown in the rotating BTZ case~\cite{Banados:1992gq}.

Explicitly,  EOM  for the above ansatz are given by
\bea 0 &=&  E_{\phi} =\frac{1}{L^2}\, e^{-2B}\Big[\Big(A'-B' +
\frac{1}{r}\Big) \phi' +\phi''\Big] - \frac{\p V}{\p
\phi}\,,\label{EOM} \\
0&=& -E_{rr} = L^2e^{2 B} V +  \frac{1}{2}\phi'^2 +  A'' +  A'^2-
A'B'- \frac{1}{r}  B'   - \frac{r^2}{2}C'^2e^{2C-2A}\,,      \nn \\
0&=& -\frac{1}{r^2} e^{2B} E_{\theta \theta} = L^2e^{2B} V + \frac{1}{r}  (A'-B') + \frac{r^2}{2 }C'^2 e^{2C-2A}\,,\nn \\
0&=& -\frac{1}{r^{2}}e^{2B-C}E_{t \theta} =  L^2 e^{2B} V   + \frac{1}{2}  \bigg[ C'' +C'^2  +r^2 C'^2e^{2C-2A}  -(A'+B')C' \nn \\
&& \qquad \qquad \qquad \qquad \qquad \qquad \qquad +\frac{2}{r}(A'-B') + \frac{3}{r}C' \bigg]\,,   \nn \\
0&=&-e^{2B} E_{tt} = L^2(r^2e^{2C} - e^{2A})e^{2B} V - e^{2A} \Big[ A''+A'^2 -A'B' +\frac{1}{r}A'\Big] \nn \\
  && \qquad \qquad \qquad + r^2e^{2C}\Big[C'' + \frac{3}{2}C'^2 + \frac{r^2}{2}C'^2e^{2C-2A} -(A'+B')C' + \frac{1}{r}(A'-B'+3C')\Big]\,, \nn
\eea
where $'$ denotes  the differentiation with respect to the radial
coordinate $r$.   These equations are called the full EOM in the following.

In Ref.~\cite{Kwon:2012zh}   extremally rotating black hole solutions with a scalar hair were found as solutions of the above EOM. It has been known that extremal BTZ black hole solutions preserve partial supersymmetry in the context of supergravity. Since the extremal black hole solutions with scalar hair can be considered as a deformation of extremal BTZ,  it is natural to expect that the supersymmetry-like argument  might play some roles to the solutions.

\section{Fake Supersymmetry and Effective Action}

 In this section, by  using the, so-called, fake supersymmetry technique, we obtain  Bogomol'nyi type of first order differential
equations which solve the full EOM.  This can be considered as the generalization of  the domain wall case to the extremally rotating $AdS_3$ black holes. This turns out to be the systematic derivation of  the first order equations for extremal black holes~\cite{Kwon:2012zh}.  It turns out that fake Killing spinor equations are not sufficient to obtain all of the first order equations. As in the case of genuine supersymmetric theory with null Killing spinors, the fake Killing spinors turn out to be null-like and should be augmented by a certain component of EOMs. In our case, by using effective action method, we show that this component of EOMs becomes effectively a first order equation and, in fact, it corresponds to a certain constraint not the dynamical equation.

\subsection{Fake supersymmetry}
 Our convention for $\Gamma$-matrices is taken such as $\{\Gamma_{\hat{a}}\, \Gamma_{\hat{b}}\} = 2\eta_{\hat{a}\hat{b}}$. Explicitly, $1+2$ dimensional (lower indices) $\Gamma$-matrices may be taken as real and symmetric ones:
\[ \Gamma^{\hat{a}}_{\alpha\beta}=(-{\bf 1}, \sigma^1, \sigma^3) \,,   \]
where $\sigma^{a}$'s are Pauli matrices.   Note that $\epsilon^{\alpha\beta}\Gamma^{\hat{a}}_{\beta\alpha}=0$.
Spinor indices are raised or lowered by rank two $\epsilon$-tensor  as
\[ \Gamma^{\hat{a}\, \beta}_{\alpha} \equiv \epsilon^{\beta\rho}\Gamma^{\hat{a}}_{\alpha\rho} =(i\sigma^2, \sigma^3, -\sigma^1)\,. \]
Then, Clifford algebra is realized as
\[ \{\Gamma^{\hat{a}}, \Gamma^{\hat{b}}\}_{\alpha}^{~\beta}  =  (\Gamma^{\hat{a}})^{~\rho}_{\alpha} (\Gamma^{\hat{b}})^{~ \beta}_{\rho} - ( \Gamma^{\hat{b}})^{~\rho}_{\alpha}(\Gamma^{\hat{a}})^{~\,\beta}_{\rho}= 2\eta^{\hat{a}\hat{b}}\delta^{~\,\beta}_{\alpha}\,. \]
We also take $\epsilon^{\hat{t}\hat{r}\hat{\theta}}=1$ such that
\[  \qquad \Gamma^{\hat{a}\hat{b}} \equiv \half [\Gamma^{\hat{a}},\Gamma^{\hat{b}}] = \epsilon^{\hat{a}\hat{b}\hat{c}}\Gamma_{\hat{c}}\,,  \qquad   \Gamma^{\hat{t}\hat{r}\hat{\theta}} =1 \,. \]
 Though there is another inequivalent irreducible representation of $\Gamma$-matrices in three dimensions, one may deal with  the inequivalent ones simply by taking  $\tilde{ \Gamma}^{\hat{a}}  \equiv -\Gamma^{\hat{a}} $.

In our case, the fake Killing spinors  under `fake' supersymmetry are determined by two equations, one of which corresponds to the (fake) dilatino variation and the other  to the (fake) gravitino variation as
\beq \Big(\Gamma^{\mu}\p_{\mu}\phi + \frac{1}{L} \frac{\p \CW}{\p \phi} \Big) \epsilon =0\,, \qquad \Big(D_{\mu} - \frac{1}{4L}\CW\Gamma_{\mu}\Big)\epsilon =0\,, \label{fake}\eeq
where  $\CW = \CW(\phi)$, the so-called
superpotential,  denotes a certain function of the scalar field $\phi$ and the curved index $\Gamma$-matrices are defined as $\Gamma^{\mu} \equiv e^{\mu}_{\hat{a}}\Gamma^{\hat{a}}$.   The  covariant derivatives in the above fake Killing spinor equations(KSE) are defined   by
\[
 D_{\mu} \epsilon \equiv \Big(\p_{\mu} + \frac{1}{4}\omega^{\hat{a}\hat{b}}_{\mu}\Gamma_{\hat{a}\hat{b}}\Big)\epsilon\,,
\]
where $\omega^{\hat{a}\hat{b}}_{\mu}$ denotes the spin connection.

The integrability conditions of the above fake KSE,   after the contraction with a $\Gamma$-matrix,  lead to the following conditions
\begin{align}
 0=&\Gamma^{\nu}\Big[ D_{\nu} - \frac{1}{4L}\CW\Gamma_{\nu},~ D_{\mu} - \frac{1}{4L}\CW\Gamma_{\mu}\Big] \epsilon = \frac{1}{2} E_{\mu\nu} \Gamma^{\nu}\epsilon\,, \label{Integrablity}\\
0=& \Gamma^{\mu}\Big[ D_{\mu} - \frac{1}{4L}\CW\Gamma_{\mu},~ \Gamma^{\nu}\p_{\nu}\phi  + \frac{1}{L}\p_{\phi}\CW\Big] \epsilon =  E_{\phi}\epsilon\,, \nn
\end{align}
where $\p_{\phi}$ denotes the differentiation with respect to the scalar field, $\phi$, and the scalar potential $V(\phi)$ should be taken in the form of
\beq V(\phi) = \frac{1}{2L^2} (\p_{\phi} \CW)^2
- \frac{1}{2L^2} \CW^2\,. \eeq
The above contracted integrability conditions show  us that EOMs for metric and scalar fields  are almost satisfied. However, as in the case of genuine Killing spinors,  the fake KSE or their integrability conditions may not   imply  the full EOM.  According to the nature of fake Killing spinors,  one may need an additional condition to imply the full EOM as will be shown in the following.

Now, let us solve the fake KSE explicitly.
For our  metric ansatz,   dreibeins  can be taken     as
\beq
 e^{\hat{t}} = L e^{A(r)} dt\,, \qquad e^{\hat{r}} = L e^{B(r)} dr\,, \qquad e^{\hat{\theta}} = L r\Big(d\theta+ e^{C(r)}dt\Big)\,.
\eeq
The spin connection one forms, $\omega^{ \hat{a}\hat{b}}=\omega_\mu^{ ~\hat{a}\hat{b}}dx^\mu $, for these dreibeins  are given by
\begin{align}
    \omega^{\hat{t}\hat{r}} &= \Big(A' e^{A-B}-\frac{1}{2}r^{2} C' e^{2C-A-B}\Big)dt -\frac{1}{2} r^{2} C'e^{C-A-B}d\theta \nn\\
   \omega^{\hat{t}\hat{\theta}} &= -\frac{1}{2} r \, C' e^{C-A}dr \\
    \omega^{\hat{r}\hat{\theta}} &= -\frac{1}{2} (2+r\,C')e^{C-B}dt -e^{-B} d\theta\,. \nonumber
\end{align}

Firstly, let us solve the fake dilatino equation.
Since the scalar field   depends only on the radial coordinate $r$ in our case, one can see that
\[
\Big(e^{-B}\phi' \Gamma^{\hat{r}} + \p_{\phi}\CW\Big)\epsilon=0 \,,
\]
which leads to
\beq
\label{KSEphi}
\Gamma^{\hat{r}} \epsilon =  \pm  \, \epsilon \,, \qquad  \phi'    =  \mp\,  e^{B} \p_{\phi}\CW \,.
\eeq
For definiteness, let us take $\Gamma^{\hat{r}} \epsilon =  \epsilon$ case, which  may be regarded as  a  projection.
By solving  directly the KSE corresponding to the fake gravitino variation,  it turns out that  the  fake Killing spinor  is a function only of the  radial coordinate $r$ and given in terms of the metric function $A(r)$ as
\beq   \epsilon_{\alpha} =   e^{A/2} \epsilon_0  \,,  \qquad
\epsilon_0 =\left( \ba{c}  1 \\  0  \ea\right)\,.\label{KS} \eeq
 Furthermore,  it turns out that metric functions and the scalar field $\phi$ should be  related through first order differential equations  as
\beq
\label{KSEpsi}
(e^{C})' = \frac{1}{r} e^{A}\Big( e^{B}\CW -
\frac{2}{r}\Big)   =   \Big( \frac{1}{r} e^{A} \Big)' \,.
\eeq
 It has been known that the KSE  imply the full bosonic EOM  if the Killing vector formed by genuine Killing spinors is time-like, while it doesn't  if the corresponding Killing vector is null-like.
 It is natural to expect the same behavior for the fake KSE. We show that in our case the Killing vector constructed from the fake Killing spinors is null-like and therefore the KSE  are insufficient to satisfy the  full EOM.

Through the standard procedure, one can construct  the one-form dual to  Killing vector by the bilinear of the fake Killing spinors as\footnote{The Dirac conjugate of spinor is defined as $\bar{\epsilon} \equiv \epsilon^{\dagger}\Gamma^{\hat{t}}$.}
\beq
 \xi \equiv \xi_{\mu} dx^{\mu} = (\bar{\epsilon} \Gamma_{\mu} \epsilon)\, dx^{\mu}\,.
\eeq
It is straightforward to check that  $\xi_{\mu}$ satisfies $\nabla_{(\mu}\xi_{\nu)} =0$ by using fake KSE, which tells us that $\xi^{\mu}$ is a  Killing vector.   Using the Fierz identity of three-dimensional $\Gamma$-matrices, it is also straightforward to see that
\beq \xi^{\mu}\xi_{\mu} =-3(\bar{\epsilon}\epsilon)^2 = 0\,, \eeq
which shows us that the Killing vector is null-like and the fake KSE is insufficient to imply full EOM. This manifests from three equations  in Eq.(\ref{KSEphi}) and (\ref{KSEpsi})  from KSE for four unknown variables.

Following the standard  way in the genuine KSE, let us identify the missing equation  for KSE to imply the full EOM in our case. To achieve this, it is convenient to introduce null coordinates adapted to the above  Killing vector as
\beq   \xi =f\, e^{\hat{+}}\,, \eeq
where $f$ is a certain normalization function.  By direct computation from the fake Killing spinor expression given in Eq.({\ref{KS}), one can take $e^{\hat{+}}$ (with $f\sim e^{A}$) as
\beq e^{\hat{+}}  \equiv  \frac{L}{\sqrt{2}} \Big[ ( r e^{C} - e^{A})dt + r d\theta\Big]\,.  \label{Nullbasis}\eeq
Then, our metric can be written as
\beq ds^2 = 2e^{\hat{+}}e^{\hat{-}} + e^{\hat{r}} e^{\hat{r}}\,, \eeq
where
\[
           e^{\hat{-}}  \equiv   \frac{L}{\sqrt{2}}\Big[ ( r e^{C} + e^{A})dt + r d\theta\Big]\,. \]
It is interesting to note that the projection condition, $\Gamma^{\hat{r}}\epsilon = \epsilon$, for fake Killing spinor implies $\Gamma^{\hat{+}}\epsilon = 0$.

Now, let us identify  the missing equation.  The following procedure  is a direct adaptation of  the  genuine Killing spinor case~\cite{Gauntlett:2002nw}\cite{Bellorin:2005hy} to the fake one.  By  the spinor contraction of $\bar{\epsilon}$ with  the contracted integrability condition, $0 = E_{\mu\nu}\Gamma^{\nu}\epsilon$,  one obtains
\beq  E_{\mu\nu}\xi^{\nu}=0 \,. \eeq
Since   $\xi^{\hat{-}}$ is   the only non-vanishing component of a Killing vector  $\xi=\xi_{\hat{+}}e^{\hat{+}}$ , the above condition implies that all the components $E_{\hat{-}\mu}$ should vanish. By multiplying   $E_{\rho\sigma}\Gamma^{\sigma}$ to the contracted integrability condition, $0 = E_{\mu\nu}\Gamma^{\nu}\epsilon$, and symmetrizing the free indices,  one  also obtains
\beq E_{\mu\rho}E_{\nu\sigma}g^{\rho\sigma}=0\,. \eeq
Using this condition (or its flat space index form), one can see that all the components of EOM are implied by fake KSE except $0=E_{\hat{+}\hat{+}}$}.
Therefore,  to imply full EOM,
 fake KSE  should be augmented by the equation $0=E_{\hat{+}\hat{+}}$, which can be written in our case as
\[ 0 =   E_{\hat{+}\hat{+}} =  \frac{ e^{-2A} }{2L^2}\Big[ E_{tt} - \frac{2}{r}( re^{C} + e^A )\, E_{t\theta} +  \frac{1}{r^2}(re^{C} +e^{A})^2E_{\theta\theta}\Big]\,. \]
Using the conditions from KSE or  the automatically vanishing components of bosonic equations, the necessary condition to imply the full EOM is given by
\beq \label{Constraint}  0 =
  E_{\hat{+}\hat{+}} =  \frac{2}{r L^{2}}\Big[  A' +B' -\frac{r}{2} \phi'^{2} \Big]\,. \eeq
In the following we will show that  this missing equation can be identified as a certain constraint not a dynamical equation in the effective action formulation.

Collecting the previous results  for   fake Killing spinors given in  Eq.~(\ref{KSEphi}) and  Eq.~(\ref{KSEpsi}) with the condition Eq.~(\ref{Constraint}), one obtains
 the following  first order differential equations, which satisfy the  full
EOM,
\bea    \label{FirstDQ}
\phi'  =   -   e^{B}  \p_{\phi} \CW\,,     \quad
  A'  + \frac{1}{r} =  e^{B}\CW\,, \quad (e^{C})'  =  \Big(\frac{1}{r}\, e^{A}\Big)' \,, \quad
A' + B' = \frac{r}{2}\phi'^2 \,.  
\eea
These differential equations,  called  reduced
EOM, were obtained by some educated guess in  Ref.~\cite{Kwon:2012zh}.

Some comments are in order. If we choose the other inequivalent  representation for $\Gamma$-matrices, $\tilde{\Gamma}$, and  take  the projection choice of fake Killing spinor as $\tilde{\Gamma}^{\hat{r}}\epsilon = \epsilon$, we obtain the same equations  in Eq. (\ref{FirstDQ}) except for the third one which changes into 
\bea
(e^{C})'  =  -\Big(\frac{1}{r}\, e^{A}\Big)' \,.
\eea
 Since the above equations  in Eq. (\ref{FirstDQ}) was derived by solving KSE for the fixed representation of $\Gamma$-matrices with definite projection $\Gamma^{\hat{r}}\epsilon = \epsilon$,  one may say that solutions of these reduced EOM preserves $1/4$ fake supersymmetries just like extremal rotating BTZ black holes\footnote{In Ref.~\cite{Colgain:2010rg}  fake supersymmetry is anticipated to play some roles  even for three-dimensional supergravity, which might be related to our case.}.  Note that the third equation in Eq. (\ref{FirstDQ}) can be integrated as
\beq e^{C} = C_{+}  + \frac{1}{r}e^{A} \,,  \eeq
where the integration constants $C_{+}$ can take any value
consistently with the  asymptotic boundary conditions.  One of the convenient choices may be to take the integration constant as
$C_{+} =0$, so that the metric function $C(r)$ is simply given by
\beq e^C =  \frac{1}{r}  e^A\,. \eeq
Note   that  the standard choice, $C_{+} =
- 1$, for instance for BTZ black holes in AdS-Schwarzschild coordinates,   can be recovered by a simple coordinate transformation,
$\theta \rightarrow \theta + C_{+}t$.
One advantage of this choice is the fact that one of the null coordinates can be identified with $\theta$ coordinate.  One can see that the Killing one-form, $\xi$, from fake Killing spinor becomes identified with $rd\theta$ as can be shown from Eq.(\ref{Nullbasis}).   This explains partially the result that the equation, $E_{\theta\theta}=0$, can be taken instead of the missing equation, $E_{\hat{+}\hat{+}}=0$ .
%

\subsection{Effective Action}
In order to clarify the nature of the missing equation in the fake Killing spinor formalism, we consider the effective action.
By inserting the metric ansatz into the action~(\ref{action}), one obtains the effective action as
\bea
S_{eff}  &=& -\frac{1}{16\pi G}\int d^3x~ L r\, e^{A-B}\Big[ 2A'' + 2A'^2 -2A'B' \\
           &&\qquad \qquad \qquad ~~~~~ ~~~~~ ~~~ - \frac{r^2}{2}C'^2\, e^{2C-2A} + \frac{2}{r}(A'-B') + \frac{1}{2}\phi'^2 + L^2e^{2B}V(\phi)\Big]   \nn \\
          & =&  -\frac{1}{16\pi G}\int d^3x~ Lr \, e^{A-B} \Big[- \frac{2}{r} A' \, - \frac{r^2}{2}C'^2\, e^{2C-2A} + \frac{1}{2}\phi'^2 + L^2 e^{2B}V(\phi)\Big] +~ \textrm{total deriv.}    \nn
 \eea
 whose  EOM  can be obtained, after rearranging results from the variation of the action with respect to $ A,\, B,\, C,\, \phi$, as
\bea
0&=& L^{2} e^{2B} V +  \frac{1}{r}(A'-B') + \frac{r^{2}}{2} C'^{2} e^{2C-2A}\,,  \\
0&=&     A'+B'- \frac{r}{2 }\phi'^{2}\,, \nn \\
0&=&  \left( r^{3} C'e^{-A-B+C} \right)'\,,  \nn \\
0&=&  L^{2} r e^{A+B}\p_{\phi}V     -\left(  r e^{A-B} \phi' \right)'  \,. \nn
\eea
One can verify that these equations are equivalent to the full EOM in Eq.~(\ref{EOM}). First of all, one may  notice that there are just four equations rather than five compared with the original full EOM. However, one can see that one of the five equations in the full EOM is redundant  as follows. Basically, the redundant equation  corresponds to the one  containing $A''$ term, for instance $0=E_{rr}$ in Eq.~(\ref{EOM}). Let us  derive this equation from the above four equations.
By differentiating  the first  equation with respect to the radial coordinate $r$,  one can obtain an equation containing $A''$ term. Though this equation also has $V'$ term,  this term  can be eliminated through the equation obtained by multiplying the last equation by $\phi'$.  By combining  the  resultant equation with the second and  third equations in the above, one can derive a differential equation containing $A''$ term which can be shown to be equivalent to $0=E_{rr}$.

Up to total derivative, the effective action can be rewritten as 
\beq
  S_{eff}  = \frac{1}{16\pi G}\int d^3x~  L \,r\, \bigg[ \frac{1}{2} r^{2} C'^{2} e^{2C-A-B}  -\frac{1}{2} e^{A+B}   \Big((\p_{\phi}\CW)^{2}-\CW^{2}\Big)     + e^{A-B}  \Big(\frac{1}{r} (A'+B') -\frac{1}{2}\phi'^{2}\Big) \bigg]\,,
 \eeq
in which it is clear that $e^{A-B}$ becomes a Lagrange multiplier and thus a variation with respect to this gives us a constraint equation,  $A'+B' - r\phi'^2/2=0$.  This equation is nothing but the missing equation obtained in Eq.~(\ref{Constraint}).  In appendix A we present preliminary study on the canonical formulation of our model to investigate the origin of this constraint.  

Let us try to extremize   the above effective action by a complete square to obtain BPS like first order equations.  By squaring the Lagrangian successively, one obtains
 \bea
 S_{eff}
   &=&  -\frac{1}{16\pi G}\int d^3x~ \frac{ Lr}{2} \bigg[  e^{A-B} \Big\{ (\phi' + e^{B} \p_\phi \CW )^{2}  -\Big(A' + \frac{1}{r}   - e^{B} \CW\Big)^{2} \Big\}   \qquad \qquad  \\
  && \qquad \qquad \qquad~~~~~   - e^{-(A+B)}\, r^{2}\Big( (e^{C})'+ (\frac{1}{r} e^{A})' \Big) \Big( (e^{C})' -  (\frac{1}{r} e^{A})' \Big)  ~\bigg]
  +\textrm{total deriv.}   \nn
\eea
One can see that the following conditions  extremize the effective action partially
\beq   \phi'  = -  e^{B} \p_\phi \CW\,, \qquad   A' + \frac{1}{r}  = e^{B} \CW\,,  \label{Partial}\eeq
which should be augmented by the constraint from Lagrange multiplier $e^{A-B}$.  After inserting the above first order equations~(\ref{Partial}) in the effective action with the constraint\footnote{The necessity of a certain constraint was noticed in a different context~\cite{HoyosBadajoz:2008fw}} ,  the  effective action  can  be further reduced as
\beq S_{eff}
   =  \frac{1}{16\pi G}\int d^3x~ \frac{ Lr}{2} e^{-(A+B)} \bigg[
      \Big( r(e^{C})' +   e^{A}(e^B\CW -\frac{2}{r}) \Big)\Big( r(e^{C})'-   e^{A}(e^B\CW - \frac{2}{r}) \Big)  ~\bigg]    +\textrm{total deriv.}
\eeq
This reduced effective action can be extremized by the following first order equation
\beq (e^{c})' = \pm \Big(\frac{1}{r}e^{A}\Big)'\,. \eeq
These seem to suggest that the effective action formalism may reproduce the first order equations obtained from fake SUSY formalism.

\section{Boundary Stress Tensor and Conserved Charges}
It was shown  that  first order differential equations derived in the previous section describe extremally rotating AdS black holes by near horizon analysis and, moreover, some of analytic solutions for these first order equations, called reduced EOM, were also presented  in Ref.~\cite{Kwon:2012zh}. In this section we obtain renormalized boundary stress tensor on the AdS black hole solutions for these reduced EOM, which is interpreted as the stress tensor of dual CFT on the asymptotic boundary by the standard AdS/CFT dictionary \cite{Balasubramanian:1999re}.  We also
confirm the extremality of these black hole solutions by  obtaining mass and angular momentum  through renormalized boundary stress tensor.
It is interesting to note that mass and angular momentum from the renormalized boundary stress tensor have contribution from both metric and scalar fields, while  two contributions are obtained in one stroke through metric  in  the so-called ADT formalism~\cite{Abbott:1981ff}\cite{Deser:2002rt}\cite{Deser:2002jk}.

The (holographically) renormalized boundary stress tensor is given by the subtraction  of  an appropriate counter term from quasi-local stress tensor introduced by Brown and York~\cite{Brown:1992br}~\cite{Balasubramanian:1999re}.    This boundary stress tensor becomes finite after the subtraction and  can be identified with the (renormalized) stress tensor in the dual field theory according to the AdS/CFT correspondence.   Using these renormalized boundary stress tensor, one can compute conserved charges in dual field theory which can also be identified with those in the bulk gravity.  In the following, we obtain the renormalized boundary stress tensor for our model and also verify the previous expressions of conserved charges. The aim of this section is two-fold.
On the one hand  we would like to obtain the contribution of a scalar hair to the boundary stress tensor and on the other we verify the conserved charge expression of our concerned black hole solutions in another way  and confirm the extremality of those black holes.

Solving reduced EOM in Eq.~(\ref{FirstDQ}) perturbatively at the asymptotic infinity, one can see that the asymptotic fall-off behaviors of AdS black hole solutions  are given by
\bea
 e^{A(r)} &=& r  + \frac{a_{1} }{r} + \cdots \,, \qquad
 e^{B(r)}  = \frac{1}{r} +\frac{b_{1} }{r^{3}} + \cdots\,, \qquad e^{C(r)}  =  -1 + \frac{1}{r}e^{A}\,,    \\
 \phi(r)   & =&\phi_{\infty} +  \frac{\phi_{1}}{r} +   \cdots\,,     \quad   \CW(\phi) = 2 + \frac{1}{2} (\phi - \phi_\infty)^2 + \cdots\,, \nn \\  \nn
\eea
where     constants $a_1$, $b_1$ and  $\phi_1$ are related as $a_1+ b_1 = -\phi^2_1/4$.  Note that  the integration constant  are taken as $C_{+}=-1$, which is more appropriate to obtain conserved charges correctly. For the superpotential $\CW(\phi)$ which is an even function of  $(\phi - \phi_\infty)$, one can show that the asymptotic form of the scalar field $\phi$ is given by
 $$  \phi(r)   =\phi_{\infty} +  \frac{\phi_{1}}{r}+ \CO\Big(\frac{1}{r^3} \Big)~. $$

In fact, by  using reduced EOM one can show that the coefficients in the next leading term is given by~\cite{Kwon:2012zh}
\beq a_1 = - \frac{1}{2}\Delta_0 \,, \quad b_1 = - \frac{1}{4}\phi^2_1+
\frac{1}{2}\Delta_0\,,
\eeq
where $\Delta_0$ is a constant related to the horizon value of the superpotential as $\Delta_0 = r^2_H\CW(\phi_H)$.  As was mentioned in the previous section, these asymptotic boundary conditions for metric functions satisfy the so-called Brown-Henneaux boundary conditions~\cite{Brown:1986nw}. Together with this metric fall-off boundary condition, the scalar field  should satisfy the similar fall-off boundary condition to be consistent with the EOM.  As an explicit example, by turning off the scalar field, that is to say, setting $\phi = \phi_{\infty}$, one obtains the extremal BTZ black holes, of which solutions are given in the above  coordinates  as
\beq
e^{A(r)}=e^{-B(r)} = r - \frac{r^2_H}{r}\,, \qquad e^{C}=
  - \frac{r^2_H}{r^2}\,, \qquad \phi = \phi_{\infty}=\phi_H\,, \qquad \CW = 2\,.  \eeq

For the boundary stress tensor computation it is very convenient to consider   the metric  foliated  in the radial direction  with  the further decomposition of the boundary metric in the ADM  form. Note that our metric ansatz is already in such a form. Explicitly,   our metric ansatz can be written as
\[
 ds^2 = N^2dr^2 + \gamma_{ij} dx^{i} dx^{j}\,, \qquad N \equiv L\, e^{B}\,,
\]
where
\[
\gamma_{ij}dx^idx^j =   -L^2e^{2A} dt^{2} + \sigma ( d\theta + e^{C} dt)^{2}\,, \qquad \sigma \equiv L^2r^2\,.
\]

As is clear from the definition of the boundary stress tensor or its unregularized Brown-York tensor form, there are two contributions to the boundary stress tensor. One contribution comes from metric fields and the other from the scalar field.
The metric contribution to the renormalized boundary stress tensor is well-known~\cite{Balasubramanian:1999re} and given in our case  by the following form
\beq
T_G^{ij}  =\frac{1}{8\pi G} \left( K \gamma^{ij} -K^{ji} -\frac{1}{L} \gamma^{ij}\right)\,,
\eeq
where $K^{ij}$ denotes the extrinsic curvature and $K$ is its trace, $K \equiv \gamma^{ij}K_{ij}$.   Our convention for the extrinsic curvature $K_{ij}$ is
\[
 K_{ij} \equiv \frac{1}{2N}\Big[\p_r \gamma_{ij} - \nabla_{i}N_{j} - \nabla_{j}N_{i}\Big]\,,
\]
where $\nabla_i$ denotes the covariant derivative with respect to the metric $\gamma_{ij}$.  Therefore, we focus only on the scalar part in the action, in the following.
Fortunately for our purpose,
the scalar field contribution to the boundary stress tensor   was  already determined for a specific scalar potential in Ref.~\cite{Gegenberg:2003jr}. However,  the fall-off boundary conditions and the    scalar potential are   different in our case from that.  Therefore, we need to rederive the scalar contribution which is appropriate in our case.

According to the standard construction of counter terms, they are chosen to cancel the unwanted divergent part of the on-shell action. To apply this procedure,  let us consider  the variation of the scalar part in our action.   After inserting the bulk EOM in the variation of the  action, one  obtains
\beq
\delta S = -\frac{1}{16\pi G}\int d^2x  \sqrt{-\gamma}  n^{\mu}\p_{\mu}\phi\, \delta \phi\,,
\eeq
where $n^{\mu}$ denotes the unit outward normal to the hypersurface or the boundary surface\footnote{In our case its nonvanishing component is just $n^r = e^{-B}/L$ or its dual one form is given by $n=Ndr$. }. To cancel this term, one needs to introduce the variation of  counter term for the scalar field as
\bea
 \delta S_{ct}  &=& \frac{1}{16 \pi G } \int_{\p M} d^{2} x \sqrt{-\gamma}\,  n^{r} \p_{r} \phi \, \delta \phi  \\
  &=& -  \frac{1}{16 \pi G } \int_{\p M} d^{2} x  \sqrt{-\gamma} \frac{1}{r^{2} L}\left[ \phi_{1}  +  \CO\Big(\frac{1}{r^2}\Big)\right] \delta \phi_1\,. \nn
\eea
In the second equality, we have expanded the integrand in powers of $1/r$  according to the fall-off boundary conditions.

Now, let us take  the integrated version of the above variational form of  counter term as
\begin{equation}
 S_{ct} = \frac{1}{16\pi G L} \int_{\p M} d^{2} x \sqrt{-\gamma} \Big[\alpha  L( \phi \, n^{r} \partial_{r} \phi ) -\beta \phi^{2} \Big]\,, \label{counter}
\end{equation}
where $\alpha$ and $\beta$ are a certain constant and will be determined in the following.
The  variation of the above  counter term leads to
\[
 \delta S_{ct} = - \frac{1}{16\pi G} \int_{\p M} d^{2} x \sqrt{-\gamma}\frac{1}{Lr^{2}}   \left[ 2(\alpha + \beta ) \phi_{1} +\CO\Big(\frac{1}{r^2}\Big) \right] \delta \phi_1\,,
\]
which should be matched to the above variational form of the counter term. This condition determines only the combination of $\alpha$ and $\beta$ as
\beq
\alpha  +  \beta =  \frac{1}{2}\,,
\eeq
which means that the counter term may  not be unique.  This is not so strange since this ambiguity does not affect the conserved charges, as will be shown in below.   It is also useful to recall that   counter terms in higher curvature gravity, which has  additional degrees of freedom through higher curvature terms,  have such ambiguity~\cite{Kwon:2011jz}\cite{Sen:2012fc}.

One can verify that conserved charges are independent of the ambiguity explicitly as follows.
First, one may note that  the renormalized boundary stress tensor is given by the sum of metric  and scalar contributions as follows:
\beq
    T^{ij}_{B}  = T_G^{ij} + T_{\phi}^{ij} =\frac{1}{8\pi G} \left( K \gamma^{ij} -K^{ji} -\frac{1}{L} \gamma^{ij}\right)+ \frac{1}{16 \pi G L} \gamma^{ij} \left( \alpha\, L \phi \, n^{r} \partial_{r} \phi - \beta \phi^{2} \right)\,,
\eeq
with the condition $ \alpha + \beta = 1/2$. Note that the scalar contribution solely comes from the the counter term in (\ref{counter}).
Then, the conserved charges can be computed by
\begin{equation}
Q_\xi= \frac{1}{8 \pi G} \int d\theta \sqrt{\sigma}~ u_{i} \xi_{j} T_{B}^{ij}\,,
\end{equation}
where $u^{i}$ and $\xi^{j}$, defined on the boundary,  denote the time-like unit vector  normal to the  hypersurface  and  a Killing vector for  the conserved charge, respectively.

To obtain the mass of our black hole solutions, one can  take  the time-like Killing vector as $\xi= e^{A} u$ with  unit one form $u=L e^{A} dt$. Then, one can see that the metric and  scalar contributions are  given,  respectively, by
\beq
M_G  = \frac{1}{8 \pi G} \int d\theta \sqrt{\sigma}~  u_{i} \xi_{ j} T_{G}^{ij} =  \frac{2\Delta_0 - \phi^2_1}{16G} \,,  \qquad
M_\phi  = \frac{1}{8 \pi G} \int d\theta \sqrt{\sigma}~  u_{i} \xi_{j} T_{\phi}^{ij}
 =    \frac{\phi^2_1}{16G} \,,
\eeq
in which $\alpha$ and $\beta$ appear only through a combination, $\alpha+\beta=1/2$. 
The total mass of the black hole solutions is given by
\beq M = M_G + M_{\phi} = \frac{\Delta_0}{8G}\,. \eeq
By taking the space-like Killing vector for angular momentum as $\xi= L r v$ with unit one form  $v=L r( d\theta +e^{C} dt) $, one obtains metric and scalar  contribution to  angular momentum as
\beq
 J_G    = \, L\frac{\Delta_{0} }{8G} \,,
\qquad
J_\phi  =  0 \,,
\eeq
which leads to the total angular momentum as
\beq J = J_G + J_{\phi} = \,  L \frac{\Delta_0}{8G}\,. \eeq
The above results on mass and angular momentum show us that the ambiguity in the counter term is harmless. Furthermore,  the expressions of conserved charges confirm  the extremality of the considered black holes $ML=  J$, which were shown  independently by  the so-called ADT formalism~\cite{Kwon:2012zh}~\cite{Nam:2010ma}. As alluded in the above, it is crucial for  the correct conserved charge that one should keep the appropriate coordinates or the appropriate integration constant $C_{+}=-1$, which was also the case in the ADT formalism.

Though the ambiguity in counter term is not physical, one may determine the counter term completely by considering more generic fall-off boundary condition for the scalar field as
\[    \phi(r) - \phi_{\infty} =\frac{\phi_{1}}{r} +   \zeta \frac{ \phi_{1}^{2}}{r^{2}}  + \cdots\,,         \]
where
$\zeta$ is  a  constant.  
Under this generalized fall-off condition for the scalar field,
 the required counter term variation becomes
\[
 \delta S_{ct}
  =-  \frac{1}{16 \pi G } \int_{\p M} d^{2} x  \sqrt{-\gamma} \frac{1}{r^{2} L}\left[ \phi_{1}  +  \zeta \frac{4}{r} \phi^{2}_1  + \cdots \right] \delta \phi_1\,.
\]
The variation of integration ansatz is given by
\[ \delta S_{ct} = - \frac{1}{16\pi G} \int_{\p M} d^{2} x \sqrt{-\gamma}\frac{1}{Lr^{2}}   \left[ 2(\alpha + \beta ) \phi_{1} +\zeta \frac{3}{r^{3}} ( 3\alpha + 2 \beta )\phi_{1}^{2}  + \cdots \right] \delta \phi_1\,,
\]
where  we have retained $\zeta$ as a constant during the variation. Comparing the above two expressions for the variation of the  counter terms, one obtains
\beq \alpha = \frac{1}{3}\,, \qquad \beta = \frac{1}{6}\,. \eeq
Then the counter term in our case can be chosen  uniquely as the limit of such a counter term by taking $\zeta \rightarrow 0$. This phenomenon such as less ambiguity for additional fall-off tail has also analogy in higher curvature gravity, where the more general fall-off solutions determine the counter terms with less ambiguity~\cite{Kwon:2011jz}.

\section{Conclusion}
In this paper we have considered fake supersymmetry to derive first order differential equations for the rotating black hole solutions in the three-dimensional Einstein gravity with a minimally coupled  self-interacting scalar field. It turns out that the fake Killing spinor is null in the sense that it leads to the null Killing vector, so that  the fake KSE should be augmented by one of EOM, $0=E_{\hat{+}\hat{+}}$ in our convention, to imply the full EOM.  We have also shown that this additional equation can be regarded  as a certain constraint by using the effective action method.

We also computed the renormalized boundary stress tensor from which we determined the mass and the angular momentum of our black hole solutions with a scalar hair. They saturate the mass bound for the angular momentum just like the usual extremally rotating BTZ black holes. 

It is somewhat unclear how to obtain all the first order equations in the effective action formalism while the fake supersymmetry formalism may not be complete in the case of null Killing spinor.  
It would be very interesting to investigate further the nature of the missing equation in the generic context of the fake supersymmetry formalism. Our investigation suggests that it may correspond to a constraint equation.  In this context it would be nice if one can identify the missing equation through the canonical approach with light-cone foliation. 

The fake supersymmetry formalism has been  a powerful tool to study the BPS states in gravity models. Since the theory itself is not supersymmetric, the solutions of fake KSE are not guaranteed to be stable. It would be an separate issue to determine the stability of those solutions. 
It would  also be very interesting to extend the fake supersymmetry formalism to the higher derivative gravity with scalar fields.


\vskip 1cm
\centerline{\large \bf Acknowledgments}
\vskip0.5cm

{We  would
like to thank  Yongjoon Kwon, Soonkeon Nam and  Jong-Dae Park for useful discussions. This work is  supported by the National
Research Foundation(NRF) of Korea grant funded by the Korea
government(MEST) through the Center for Quantum
Spacetime(CQUeST) of
Sogang University with grant number 2005-0049409. SH is supported by the National Research Foundation of Korea(NRF) grant funded 
by the Korea government(MEST) with the grant number  2012046278.  S.H.Y is supported
  by Basic Science Research Program through the NRF of Korea funded by the MEST(2012R1A1A2004410). }
\newpage

\appendix
\centerline{\large \bf Appendix}

 \renewcommand{\theequation}{A.\arabic{equation}}
  \setcounter{equation}{0}
\section{Canonical Formalism}

In this appendix, we describe the canonical formalism of our model. Since the canonical formalism for the scalar field is trivial, we focus on the formalism for the metric. The aim of this section is to indicate that the missing equation in the fake supersymmetry formalism may be connected with the Hamlitonian and momentum constraints. Here, we adopt the standard notation in the canonical formulation with time-like foliation, which will be used only in this appendix.  

Through the ADM decomposition of the metric as
\beq ds^2 = -N^2dt^2 + \gamma_{ij}(dx^i + N^idt) (dx^j +N^jdt)\,, \eeq
one can apply the canonical formalism to gravity. In this formulation, $\gamma_{ij}$'s are taken as canonical variables and their conjugate momentums are given in terms of the extrinsic curvature $K_{ij}$ by
\beq \pi^{ij} \equiv \sqrt{\gamma}\Big[K^{ij} - \gamma^{ij} K\Big]\,. \eeq
In our convention the extrinsic curvature $K_{ij}$ is  defined by
\[
 K_{ij} \equiv \frac{1}{2N}\Big[\p_t \gamma_{ij} - \nabla_{i}N_{j} - \nabla_{j}N_{i}\Big]\,, \qquad K \equiv \gamma^{ij}K_{ij}
\]
where $\nabla_i$ denotes the covariant derivative with respect to the metric $\gamma_{ij}$.

By diffeomorphism invariance, one obtains constraints which are called as Hamiltonian and momentum constraints. These constraints can be written in our case respectively as
\bea 0 &=& {\cal H} =  -\sqrt{\gamma}\Big[{}^{(2)}R  - \frac{1}{2} \p^{i}\phi \p_{i}\phi   -V \Big] +  \frac{1}{\sqrt{\gamma}}\Big[\pi^{ij}\pi_{ij} - \pi^2 + \frac{1}{2}\pi^2_{\phi}\Big]\,, \\
0 &=& {\cal P}_i =  -2\sqrt{\gamma}~   \nabla_{j}\Big(\frac{1}{\sqrt{\gamma}}\pi^{j}_{i}\Big) + \pi_{\phi} \p_{i}\phi\,, \nn \eea
where ${}^{(2)}R$ denotes the curvature scalar in two dimenisons for $(r, \theta)$ and
\[
\pi_{\phi} \equiv  -\frac{1}{N}\sqrt{\gamma}\Big(\p_{t}\phi -N^i\p_i\phi\Big) \]
denotes the conjugate momentum for the scalar field $\phi$.
Using our antatz for the black hole metric, one can see that these constraints lead to
\bea 0 &=& L^2e^{2B}V + \frac{1}{2}\phi'^2 - \frac{2}{r}B' + \frac{r^2}{2}(e^C)'^2e^{-2A}\,,  \\
       0 &=& \Big(r^3e^{-(A+B)}\, (e^{C})' \Big)' \,. \nn
\eea

The canonical Hamiltonian is given by
\beq
 H = \int d^2x \sqrt{\gamma}\Big[N{\cal H} + N^i{\cal P}_i \Big] + {\rm surface~term}\,,
\eeq
and  dynamical equations in the canonical formalism are given by
\[
  \frac{\delta H}{\delta \pi^{ij} }=  \p_{t}\gamma_{ij}\,, \qquad \frac{\delta H}{\delta \gamma_{ij}} = - \p_{t}\pi^{ij}\,,
\]
where the first equation  is nothing but the condition determining  the extrinsic curvature  by $\p_{t}{\gamma}_{ij}$.
In terms of the extrinsic curvature $K_{ij}$,   the second dynamical equation can be written as
\bea \p_{t}K_{ij} &=&  N^{k}\nabla_{k} K_{ij} +K_{ik} \nabla_{j}N^{k} + K_{jk} \nabla_{i}K^{k}  + \nabla_{i}\nabla_{j}N    \\
                       & &   -  N\Big[{}^{(2)}R_{ij} - 2K_{i}^{k}K_{kj}   +  KK_{ij} - \frac{1}{2}\p_{i}\phi\p_{j}\phi -  \gamma_{ij}V(\phi)\Big] \,. \nn
\eea
As is clear from this expression,  this equation leads to two equations among EOM for the metric field  as follows
\bea
0&=&  L^2e^{2 B} V +  \frac{1}{2}\phi'^2 +  A'' +  A'^2-
A'B'- \frac{1}{r}  B'   - \frac{r^2}{2}C'^2e^{2C-2A}\,,  \nn     \\
0&=&   L^2e^{2B} V + \frac{1}{r}  (A'-B') + \frac{r^2}{2 }C'^2 e^{2C-2A}\,. \nn
\eea
These equations correspond  to $0=E_{rr}$ and  $0=E_{\theta\theta}$     in the full EOM.
If one use the first order equations obtained from fake supersymmetry formalism, the combination of those constraints and the equation  $0=E_{\theta\theta}$ becomes  the missing equation. 
This indicates that the constraint by the Lagrange multiplier $e^{A-B}$ in the effective action may appear in the light-cone foliation.

\newpage

\end{document}